\documentclass[%
reprint,
superscriptaddress,
amsmath,amssymb,
aps,
prb,
]{revtex4-1}
\usepackage{bm}
\usepackage{color, soul}
\usepackage{amsmath}
\usepackage{natbib}
\usepackage{bm}
\usepackage{times}
\usepackage{url}
\usepackage{amssymb}
\usepackage{braket}
\usepackage{mathtools}
\usepackage{commath}
\usepackage{bigints}
\usepackage{booktabs}
\usepackage{float}
\usepackage{multirow}
\usepackage{graphicx}% Include figure files
\usepackage{dcolumn}% Align table columns on decimal point
\usepackage{bm}% bold math
\def \sample {Cu$_2$OSO$_4$}
\usepackage{subfigure}
\DeclareUnicodeCharacter{2212}{-}

\begin{document}
	
	%\preprint{APS/123-QED}2
	
	\title{Ferrimagnetic 120$^\circ$ magnetic structure in \sample}% Force line breaks with \\
	%	\thanks{A footnote to the article title}%
	%	
	%	\author{Virgile Yves Favre}	
	%	\affiliation{Laboratory for Quantum Magnetism, Institute of Physics, Ecole Polytechnique Féderale de Lausanne, CH-1015 Lausanne, Switzerland}
	%	\email{virgile.favre@epfl.ch}
	%	
	%	
	%	\author{Ivica \v{Z}ivkovi\'c}	
	%	\affiliation{Laboratory for Quantum Magnetism, Institute of Physics, Ecole Polytechnique Féderale de Lausanne, CH-1015 Lausanne, Switzerland}
	%	\email{ivica.zivkovic@epfl.ch}
	%	
	%	
	%	\author{H. M. R\o nnow}	
	%	\affiliation{Laboratory for Quantum Magnetism, Institute of Physics, Ecole Polytechnique Féderale de Lausanne, CH-1015 Lausanne, Switzerland}
	%	\email{henrik.ronnow@epfl.ch}
	%	
	%	
	%	\author{Ann Author}
	%	\altaffiliation[Also at ]{Physics Department, XYZ University.}%Lines break automatically or can be forced with \\
	%	\author{Second Author}%
	%	\email{Second.Author@institution.edu}
	%	\affiliation{%
	%		Authors' institution and/or address\\
	%		This line break forced with \textbackslash\textbackslash
	%	}%
	%	
	%	\collaboration{MUSO Collaboration}%\noaffiliation
	%	
	%	\author{Charlie Author}
	%	\homepage{http://www.Second.institution.edu/~Charlie.Author}
	%	\affiliation{
	%		Second institution and/or address\\
	%		This line break forced% with \\
	%	}%
	%	\affiliation{fdim
	%		Third institution, the second for Charlie Author
	%	}%
	%	\author{Delta Author}
	%	\affiliation{%
	%		Authors' institution and/or address\\
	%		This line break forced with \textbackslash\textbackslash
	%	}%
	%	
	%	\collaboration{CLEO Collaboration}%\noaffiliation

	\author{Virgile Yves Favre}
	\affiliation{Laboratory for Quantum Magnetism, Institute of Physics, Ecole Polytechnique Féderale de Lausanne, CH-1015 Lausanne, Switzerland}
	\author{Gregory S. Tucker}
	\affiliation{Laboratory for Quantum Magnetism, Institute of Physics, Ecole Polytechnique Féderale de Lausanne, CH-1015 Lausanne, Switzerland}
	\affiliation{Laboratory for Neutron Scattering, Paul Scherrer Institut, CH-5232 Villigen PSI, Switzerland}
	\author{Clemens Ritter}
	\affiliation{Institute Laue Langevin, BP 156, 38042, Grenoble, France.}
	\author{Romain Sibille}
	\affiliation{Laboratory for Neutron Scattering, Paul Scherrer Institut, CH-5232 Villigen PSI, Switzerland}
	\author{Pascal Manuel}
	\affiliation{ISIS Facility, STFC Rutherford Appleton Laboratory, Oxfordshire OX11 0QX, UK}
	\author{Matthias D. Frontzek}
	\affiliation{Neutron Scattering Division, Oak Ridge National Laboratory, Oak Ridge, Tennessee 37830, USA}
	\author{Markus Kriener}
	\affiliation{RIKEN Center for Emergent Matter Science, Wako 351-0198, Japan.}
	\author{Lin Yang}
	\affiliation{Laboratory for Quantum Magnetism, Institute of Physics, Ecole Polytechnique Féderale de Lausanne, CH-1015 Lausanne, Switzerland}
	\affiliation{Laboratory of Physics of Complex Matter, Institute of Physics, Ecole Polytechnique Féderale de Lausanne, CH-1015 Lausanne, Switzerland}
	\author{Helmut Berger}
	\affiliation{Crystal Growth Facility, Ecole Polytechnique Fédérale de Lausanne, Lausanne, Switzerland.}
	\author{Arnaud Magrez}
	\affiliation{Crystal Growth Facility, Ecole Polytechnique Fédérale de Lausanne, Lausanne, Switzerland.}
	\author{Nicola P. M. Casati}
	\affiliation{Swiss Light Source, Paul Scherrer Institut, 5232, Villigen, Switzerland}
	\author{Ivica \v{Z}ivkovi\'c}
	\email{ivica.zivkovic@epfl.ch}
	\affiliation{Laboratory for Quantum Magnetism, Institute of Physics, Ecole Polytechnique Féderale de Lausanne, CH-1015 Lausanne, Switzerland}
	\author{Henrik M. R\o nnow}
	\email{henrik.ronnow@epfl.ch}
	\affiliation{Laboratory for Quantum Magnetism, Institute of Physics, Ecole Polytechnique Féderale de Lausanne, CH-1015 Lausanne, Switzerland}
	
	\date{\today}% It is always \today, today,
	%  but any date may be explicitly specified% https://en.wikipedia.org/wiki/Einstein_solid
	
	\begin{abstract}
		We report magnetic properties of a 3d$^9$ (Cu$^{2+}$) magnetic insulator \sample \ measured on both powder and single crystal. The magnetic atoms of this compound form layers, whose geometry can be described either as a system of chains coupled through dimers or as a Kagom\'e lattice where every 3rd spin is replaced by a dimer. Specific heat and DC-susceptibility show a magnetic transition at 20 K, which is also confirmed by neutron scattering. Magnetic entropy extracted from the specific heat data is  consistent with a $S=1/2$ degree of freedom per Cu$^{2+}$, and so is the effective moment extracted from DC-susceptibility. The ground state has been identified by means of neutron diffraction on both powder and single crystal and corresponds to a $\sim120$ degree spin structure in which ferromagnetic intra-dimer alignment results in a net ferrimagnetic moment. No evidence is found for a change in lattice symmetry down to 2 K. Our results suggest that \sample \ represents a new type of model lattice with frustrated interactions where interplay between magnetic order, thermal and quantum fluctuations can be explored.
		
		%		                                                                                                                       
		%		\begin{description}
		%			\item[Usage]
		%			Secondary publications and information retrieval purposes.
		%			\item[PACS numbers]
		%			May be entered using the \verb+\pacs{#1}+ command.
		%			\item[Structure]
		%			You may use the \texttt{description} environment to structure your abstract;
		%			use the optional argument of the \verb+\item+ command to give the category of each item. 
		%		\end{description}
	\end{abstract}
	\pacs{Valid PACS appear here}% PACS, the Physics and Astronomy
	% Classification Scheme.
	%\keywords{Suggested keywords}%Use showkeys class option if keyword
	%display desired
	\maketitle
	
	{%\twocolumn
		\section{Introduction}
			Materials with antiferromagnetic Heisenberg interactions between spins on a triangular lattice\cite{trian1, trian2,trian3} inherently exhibit large frustration, resulting in many similar energy states giving rise to novel behavior. In practice, there exist several ways to build a full lattice from triangular motifs. The notion of quantum spin liquid was introduced in the form of the Resonating Valence Bond, as a potential ground state of the simple triangular lattice \cite{spinliquids1,spinliquids2}system. Later developments evidenced that both classical and quantum ground state of the triangular lattice  actually displayed long range order in the so-called 120$^\circ$ configuration\cite{triangle120}.
			Among an intense search for quantum spin liquids in models and materials\cite{francesco, otto}, considerable attention has been devoted experimentally \cite{kagome1, kagome2, kagome3, kagome4}, and theoretically \cite{kagometeh1, kagometeh2, kagometeh3} to the so-called Kagom\'e lattice, which is obtained by removing 1/4 of the spins from a triangular lattice, leading a lower connectivity of four nearest neighbors instead of six, which enhances fluctuations. It has been argued theoretically that the S = 1/2 Kagom\'e Heisenberg antiferromagnet shows a spin-liquid ground state \cite{kagomeGS1, kagomeGS2, kagomeGS3} and it is still debated whether the resulting states should be gapped or not \cite{kagomeExcitations1, kagomeExcitations2, kagomeExcitations3}.\\
			 Many experimental attempts have been made to find compounds hosting this rich Kagom\'e physics \cite{kagomeTest1, kagomeTest2, kagomeTest3, kagomeTest4, kagomeTest5}. Herbertsmithite\cite{herbert} remains one of the best candidates. Many other candidates exhibited deviations from the Kagom\'e model, such as: magnetic interactions that go beyond nearest neighbor \cite{kagomeNext}; non-negligible out-of-plane coupling; a significant antisymmetric Dzyaloshinskii-Moriya (DM) interaction \cite{dm1, dm2}, structural distortions or chemical disorder. These perturbations lift the degeneracy of an ideal spin-liquid and drive the system towards long-range magnetic order.\\
			 A major question that currently has no clear answer is to which extent quantum spin liquid properties can survive and co-exist with ordered states. This question extends beyond Kagom\'e related systems\cite{DallaPiazza2015}. It is therefore interesting to study other triangle-containing lattice geometries than the pure triangular and Kagom\'e lattices.\\
	In this context we present the study of \sample\cite{actaCrystallo}, which hosts a quasi-2D lattice built from triangular motifs. One way to describe the  lattice  is that of a Kagom\'e lattice with one third of the sites replaced by $S=1/$ pairs (dimers). 
%with intra-dimer coupling, effectively holding S=1 magnetic moments. 
	Another way to characterize it would be chains coupled through spin pairs in a triangular frustrated pattern . 
	As shall be presented below, the spin-pair appear ferromagnetically aligned, such that an effective model for the system could be described as a Kagome lattice with $S=1/2$ on 2/3 of the sites and $S=1$ on 1/3 of the sites. 
	No single crystal growth of \sample has previously been reported\cite{fail}. Studies of powder samples of \sample \cite{french, vileneuve} reported the DC susceptibility of the compound and showed evidence of a transition to a magnetically long range ordered state at 20 K. The DM interaction in the sample was estimated by the means of Electron Spin Resonance (ESR) \cite{japan}. From bulk measurements, a non-colinear antiferromagnetic ground state was proposed \cite{magStructFai}. Here we report single crystal growth and the details of the ground state of \sample, derived from DC-susceptibility, magnetization, heat capacity, as well as x-ray and neutron scattering.
	
		\section{Experimental details}
		Single crystals of \sample \ were grown by chemical vapor transport. High quality \sample  \  powder was synthesized using anhydrous CuSO$_4$ as a source in a quartz crucible placed in the center of a muffle furnace and heated at 740 $^\circ$C. Two different transport agents were placed in a quartz ampoule at room temperature: Cl$_2$ and NiBr$_2$, as well as a portion of the \sample \ powder. The ampoules were placed in a two zone gradient furnace. The best charge and growth-zone temperatures were 650 $^\circ $C and 550 $^\circ$C respectively. After five weeks, several dark-brown, semitransparent crystals were obtained. The typical dimensions of the crystals are 4x2x0.5 mm$^3$. On most of the crystals the b axis can be identified as an edge and most of them also present a (001) facet.\\
		Specific heat was measured using a physical properties measurement system (PPMS, Quantum Design, Inc) and magnetization was measured using both PPMS and a magnetic properties measurement system (MPMS, Quantum Design, Inc). Neutron diffraction experiments were performed on WAND, ZEBRA, D20 and WISH beam lines at ORNL, PSI, ILL and ISIS respectively. The measurement on D20 was done using the high resolution option with takeoff angle 90$^\circ$ and a wavelength of 2.41 \AA. 4 hour measurements were taken at 1.5 K and at 30 K while the temperature ramp was done between 1.5 K and 30 K using 30 min runs with about 0.45 K between consecutive runs. Synchrotron x-ray diffraction was performed at the MS-X04SA beam line\cite{willmott2013_X04SA} at SLS, PSI. For powder diffraction, crushed single crystals were used to minimize impurities.
		\section{Results}
		\subsection{X-ray diffraction and crystal structure}
		According to a previous study \cite{actaCrystallo}, the compound belongs to the monoclinic space group C 2/m with lattice parameters a = 9.355(10) \AA, b = 6.312(5) \AA, c = 7.628(5) \AA, $\alpha$ = $\gamma$ = 90$^\circ$, and $\beta$ = 122.29$^\circ$.\\
		To determine the temperature dependence of the nuclear structure, we performed temperature dependent x-ray diffraction on a powder sample. The temperature dependence of lattice parameters was extracted from LeBail fits (LBF) and is shown in Fig. \ref{figxray:sfig4}, as well as the unit cell volume. The overall thermal contraction of the lattice has been observed down to 50 K, with a minimum of lattice parameters a and b around 60 K. This non-monotonic temperature dependence could be due to the onset of magnetic correlations coupling to the lattice. Similar magneto-elastic coupling is observed below the ordering temperature (Fig. \ref{neutron_lattice}). 
		
		\begin{figure}[htp]
			\centering
			\includegraphics[width=0.45\textwidth]{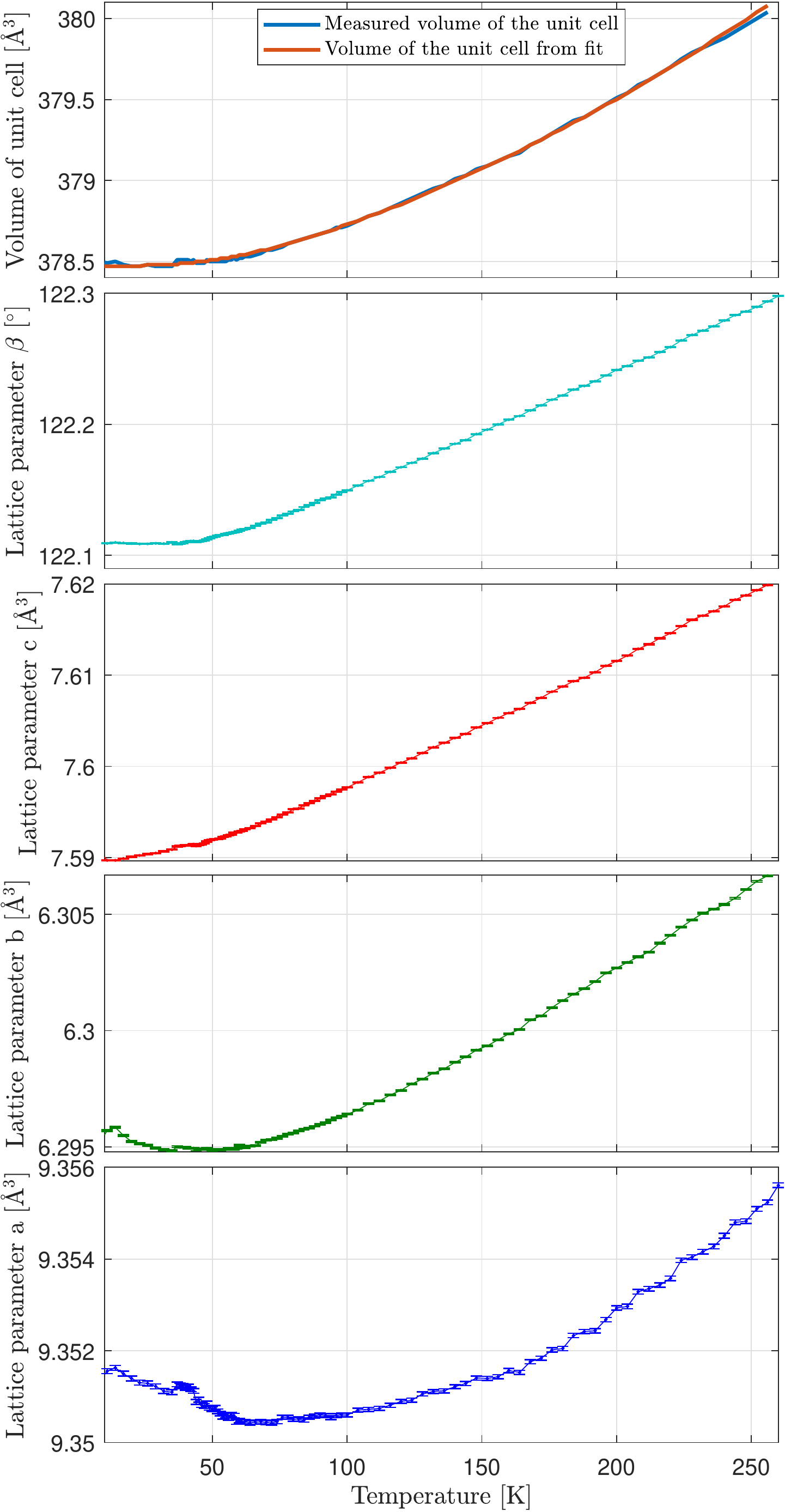}
			\caption{Change in lattice parameters and unit cell volume obtained from LeBail fitting of powder x-ray diffraction data.\label{figxray:sfig4}}
		\end{figure}
		\begin{figure}[htp]
			\centering
			\includegraphics[width=0.415\textwidth]{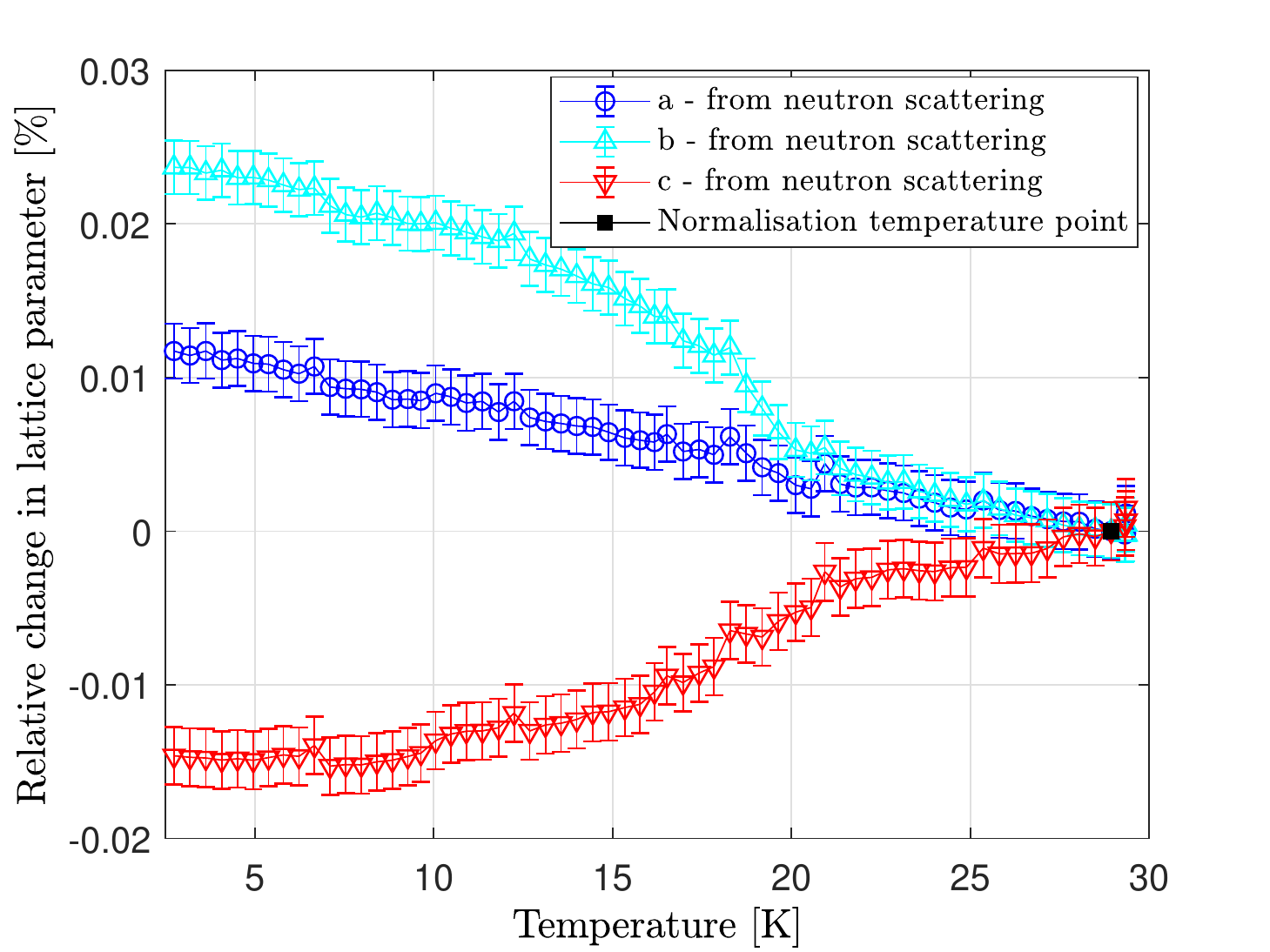}
			\caption{Relative changes in lattice parameters obtained from neutron diffraction. The data has been normalized to data measured at 29 K (black square). \label{neutron_lattice}}
		\end{figure}
		
		%\subsection{Crystal structure}
			The crystal structure of \sample \  gives rise to an interesting magnetic lattice. The copper atoms occupy two inequivalent positions. They are arranged in the ab-plane, with atoms on copper site one (Cu$_1$) exactly in the plane, while atoms of copper site two (Cu$_2$) are positioned symmetrically above and below the plane. Fig. \ref{fig:sfig3} shows how the planes are then interconnected through SO$_4$ tetrahedra \cite{vesta}.  
			The arrangement of Cu$_1$ and Cu$_2$ ions in the quasi-2D planes can be described either as Cu$_1$ chains along the crystallographic b-axis, interconnected by Cu$_2$ dimers in a frustrated zig-zag pattern; or it can be described as aKagom\'e lattice with one third of the sites replaced by dimers, or it can be viewed as Cu$_1$ chains coupled through Cu$_2$ dimers.  Fig. \ref{fig:sfig1} illustrate how viewing the magnetic layers rotated 8.5$^\circ$ around the b-axis plane overlays the Cu$_2$ ions resulting in the familiar Kagom\'e motif. Fig. \ref{ground state} show the magnetic lattice viewed directly along $c^*$.
%		\begin{figure}
%			\begin{subfigure}{.5\textwidth}
%				\centering
%				\includegraphics[width=\linewidth]{../pictures/polished/interlayer_so4_reviewed_2.jpg}
%				\caption{View from the b axis, showing a slice of the Kagom\'e-like planes formed by copper atoms in \sample, highlighting the SO$_4$ tertrahedra connecting the copper planes.}
%				\label{fig:sfig3}
%			\end{subfigure}
%			\begin{subfigure}{.5\textwidth}
%				\centering
%				\includegraphics[width=\linewidth]{../pictures/polished/kagome_plane_hr_leg_reviewed_2.jpg}
%				\caption{Kagom\'e-like planes formed by copper atoms in \sample \ in the ab-plane. The two colors correspond to two inequivalent copper sites.}
%				\label{fig:sfig1}
%			\end{subfigure}%
%			\caption{\sample  \  structure obtained from X-ray diffraction.}
%			\label{fig:fig}
%		\end{figure}
	\begin{figure}
		\centering
		\subfigure[View from the b axis, showing a slice of the Kagom\'e-like planes formed by copper atoms in \sample, highlighting the SO$_4$ tertrahedra connecting the copper planes.]{\label{fig:sfig3}\includegraphics[width=\linewidth]{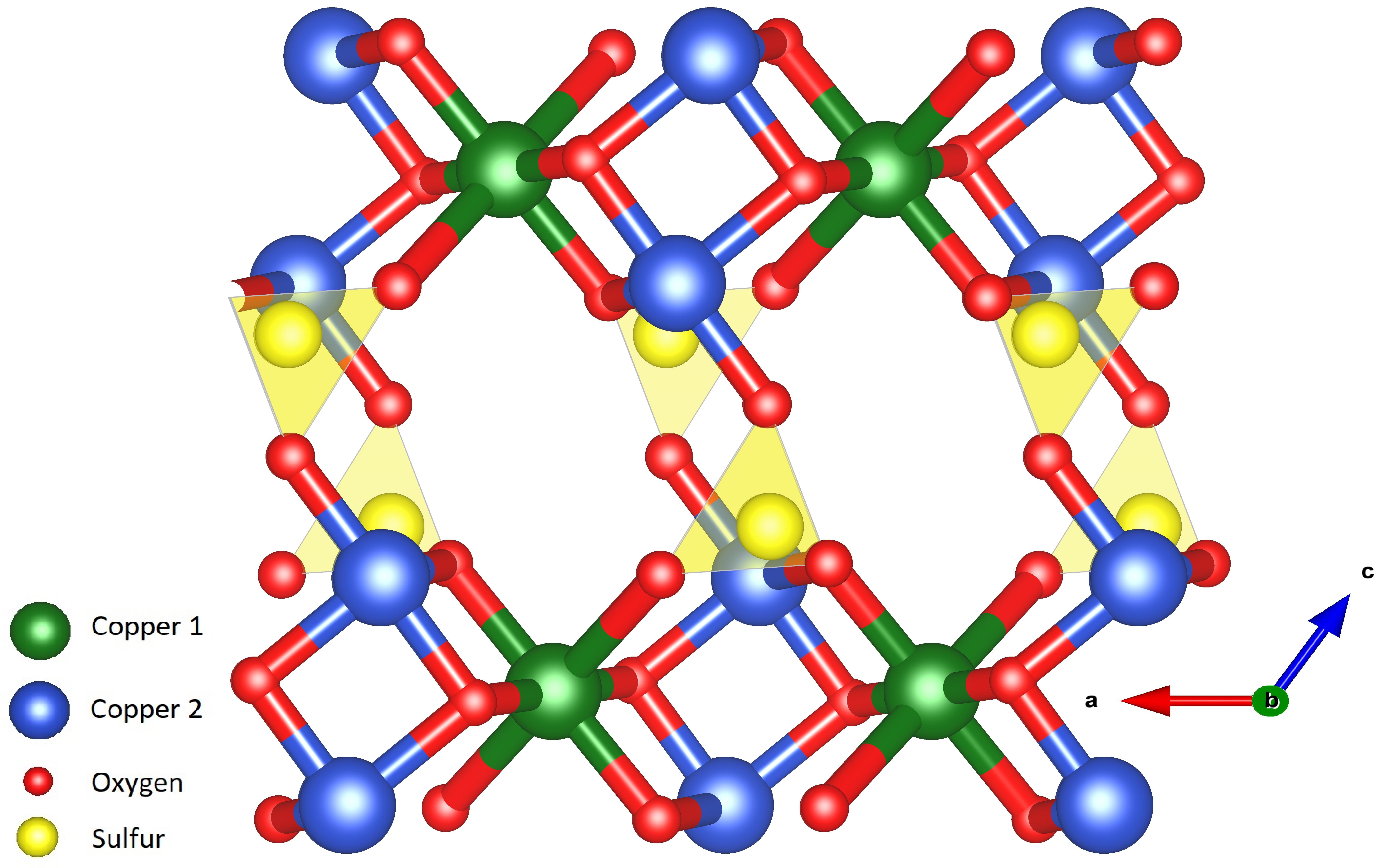}}
		\subfigure[Kagom\'e-like planes formed by copper atoms in \sample \ in the ab-plane. The two colors correspond to two inequivalent copper sites.]{\label{fig:sfig1}\includegraphics[width=\linewidth]{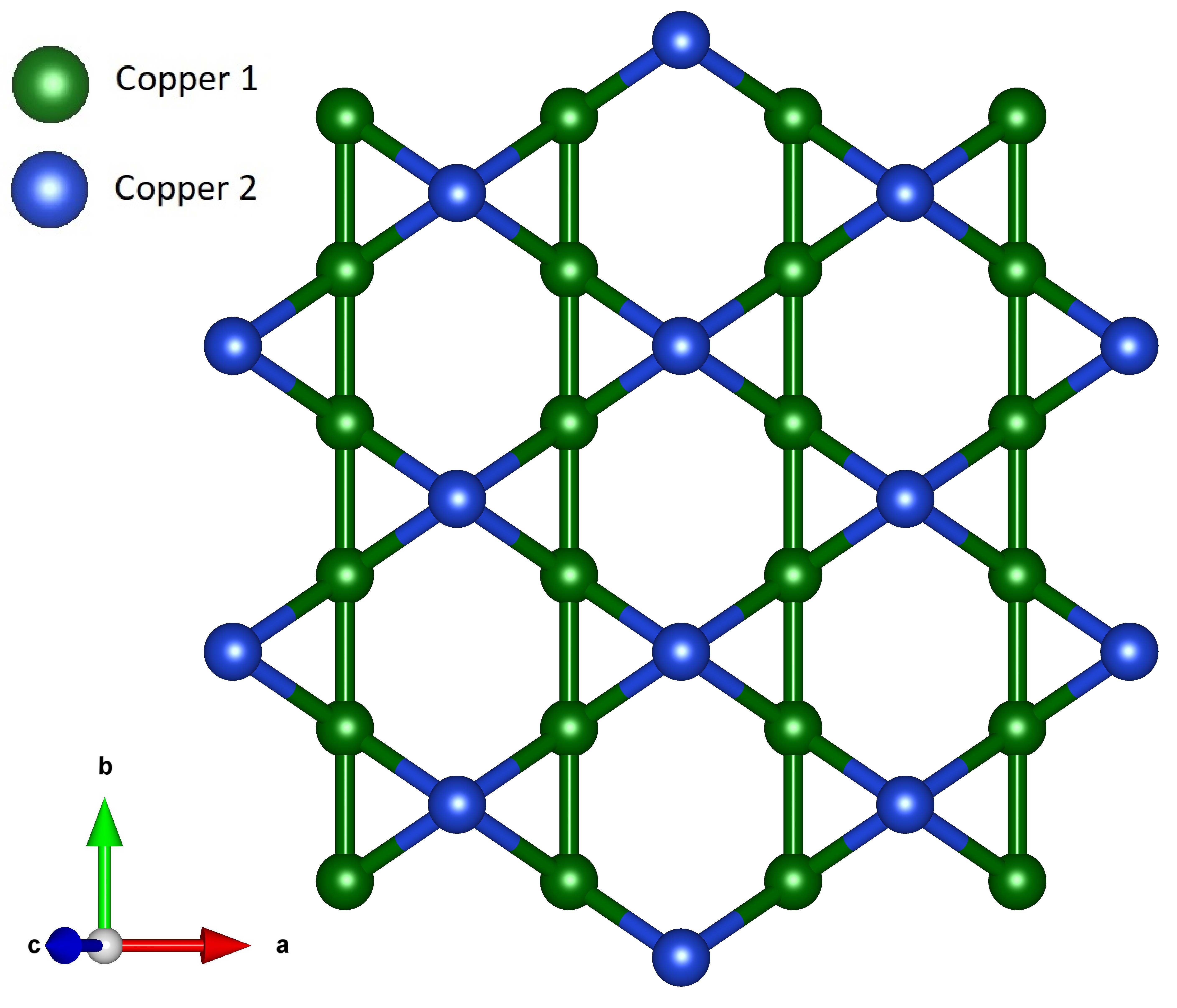}}
		\caption{\label{fig:fig} \sample  \  structure obtained from X-ray diffraction.}
	%%			\caption{View from the b axis, showing a slice of the Kagom\'e-like planes formed by copper atoms in \sample, highlighting the SO$_4$ tertrahedra connecting the copper planes.}
	%			\label{fig:sfig3}
	\end{figure}
		\subsection{Specific Heat}
		The specific heat C$_p$ measured from 2 K to 200 K in zero field and in a magnetic field of 9 T is shown in Fig. \ref{specific heat}. The C$_p$ above 70 K for both fields is essentially the same, increasing monotonically with increasing temperature. In zero field a pronounced peak is found at 20 K, corresponding to a transition into a magnetically long range-ordered phase, as evidenced by other experimental measurements discussed in later sections. The peak shifts to slightly higher temperature at 9 T.\\
		In order to extract the magnetic part of the specific heat, C$_\text{mag}$, and to deduce the corresponding entropy S$_\text{mag}$, we simulate the lattice contribution from the high temperature data by taking into account one Debye (C$_D$) and several Einstein (C$_{E,i}$) contributions. We use a combined fit to describe the C$_p$ and the volume of the unit cell obtained from x-ray diffraction by a phonon (lattice only) model. Similarly to what has been done in \cite{hc, hadi}, the lattice contribution to the specific heat is given by $C_p = C_D + \sum_i C_{E,i}$ with :
		\begin{equation} C_D = 9 n_D R \left( \frac{T}{\Theta_D} \right)^3 \int_0^{\Theta_D / T} \frac{x^4 e^x}{\left(e^x - 1 \right)^2} dx
		\end{equation}
		and
		\begin{equation}
		C_E = 3n_ER\frac{y^2e^y}{\left( e^y - 1 \right)^2}, y \equiv \Theta_E/T 
		\end{equation}
		Where R denotes the gas constant, $\Theta_D$ and $\Theta_E$ are the Debye and Einstein temperatures respectively. The sum n$_D$ + n$_E$ is the total number of atoms per formula unit. The volume of the unit cell has been fitted together with the specific heat using the Debye and Einstein contributions to the internal energy. The volume of the unit cell is related to the internal energy by \cite{prb, prb2, einsteinn}: 
		\begin{equation}
		V(T)=\gamma U(T) / K_{0}+V_{0}
		\end{equation}
		Where $V_0$ is the cell volume at T = 0 \ K, K$_0$ is the bulk modulus and $\gamma$ is the Gr\"uneisen parameter. U(T) is the internal energy which can be expressed in terms of the Debye and Einstein approximation as : 
		\begin{equation}
		U(T) = U_\mathrm{D}(T) + U_\mathrm{E}(T)
		\end{equation}
		\begin{equation}
		U_\mathrm{D}(T)=9 n_D k_{\mathrm{B}} T\left(\frac{T}{\Theta_{\mathrm{D}}}\right)^{3} \int_{0}^{\Theta_{\mathrm{D}} / T} \frac{x^{3}}{e^{x}-1} d x
		\end{equation}
		\begin{equation}
		U_\mathrm{E}(T) = \frac{3}{2}k_\mathrm{B} \sum_i n_\mathrm{E,i}\Theta_\mathrm{E,i} \coth{} (\frac{\Theta_\mathrm{E,i}}{2T})
		\end{equation}
		The best fit for both sets of data, using one Debye branch and two Einstein branches yields the characteristic temperatures : $\Theta_D$ = 171 K, $\Theta_{E1}$ = 245 K and $\Theta_{E2}$ = 939 K with n$_D$ = 1, n$_{E1}$ = 3 and n$_{E2}$ = 4. This fit was performed using data for temperatures larger than 70 K. The volume of the unit cell at zero Kelvin has also been fitted to be : $V_0$ = 373.65 \AA$^3$, (Fig. \ref{figxray:sfig4}).\\
		Fig. \ref{specific heat} show the resulting C$_\text{mag}$/T in zero field as well as the magnetic entropy, S$_\text{mag}$(T) obtained by integrating C$_\text{mag}$/T over temperature. The magnetic entropy saturates around 101(19)\% of R ln 2 per formula unit, which is in agreement with the entropy of a two level spin half system. A similar analysis carried out on the 9 T data (not shown) indicates negligible field effects. The magnetic entropy is shared between short range correlations (Fig. \ref{specific heat}, green shading) and phase transition to long range order(Fig. \ref{specific heat}, blue shading). We estimate that 18(5)\% of the magnetic entropy contributes to the phase transition.
		\begin{figure}[htp]
			\includegraphics[width=0.45\textwidth]{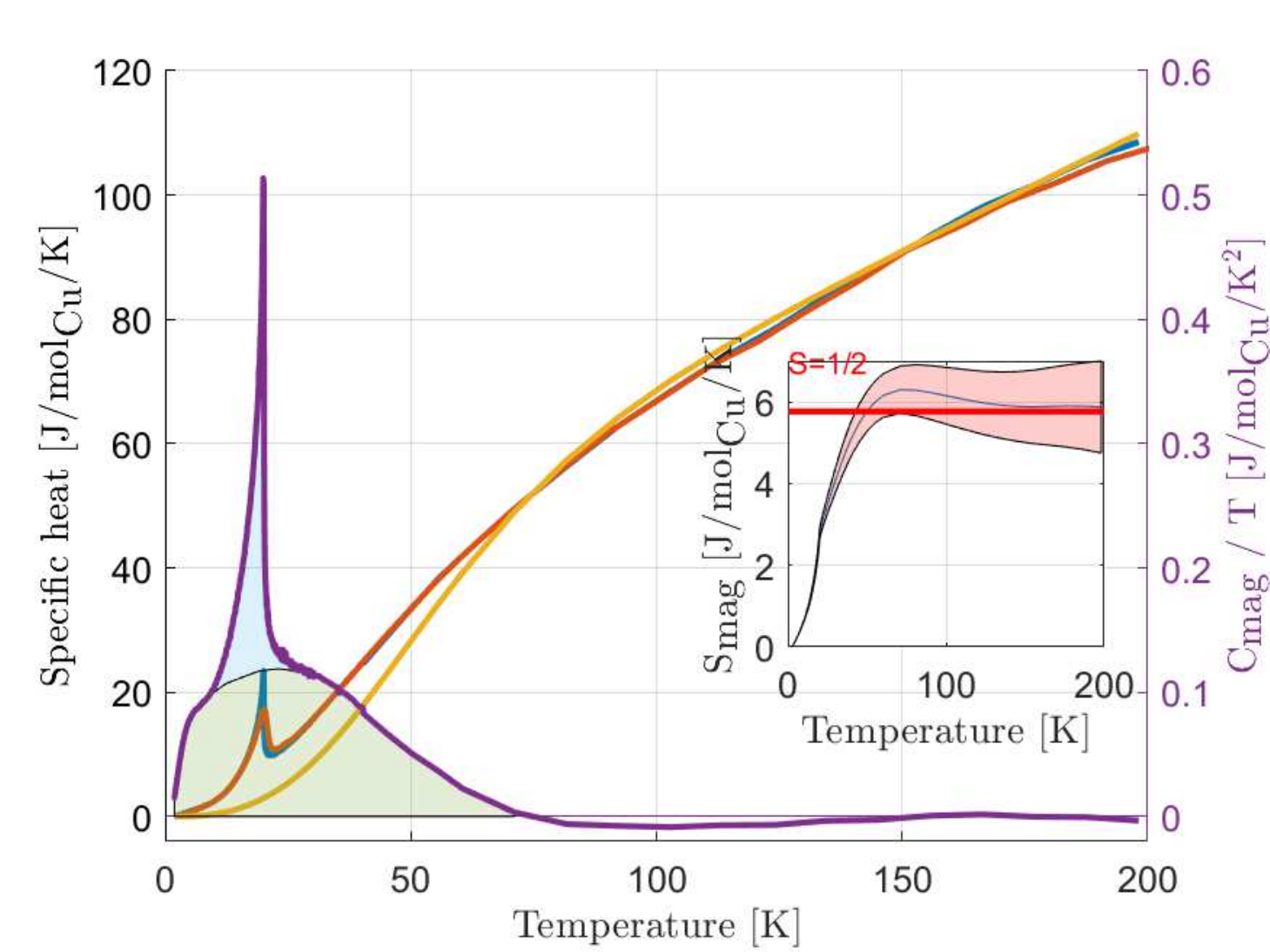}
			\caption{\label{specific heat}Specific heat C$_p$ as a function of temperature in zero field (blue) and 9T (red). Yellow line is the fitted lattice contribution. The purple curve (right vertical axis) shows the magnetic part of the specific heat, C$_{\textrm{mag}}$/T. The blue and green shadings represent the estimated C$_\textrm{mag}/$T that contributes respectively to the phase transition and to short range correlations. The inset presents the magnetic entropy, $\textrm{S}_{\textrm{mag}}(\textrm{T}) = \int \textrm{C}_{\textrm{mag}}(\textrm{T})/\textrm{T} \textrm{dT}$. Red shading shows the confidence interval.}
		\end{figure}
		
		\subsection{Susceptibility and magnetization}
		The inverse DC susceptibility measured on a single crystal, with a field of 0.1 T, pointing along the three relevant crystallographic directions : a, b, and c* is shown in Fig. \ref{cw_fits}. The high temperature part of $\chi$(T) fits the Curie-Weiss law, $\chi$(T) = C/(T-$\Theta_\text{CW}$) + $\chi_0$, where $\Theta_\text{CW}$ is the Curie-Weiss temperature, $\textrm{C} = \mu_{\textrm{eff}}^2/8$ and $\chi_0$ a temperature independent diamagnetic and background term. $\mu_{\textrm{eff}} = g \sqrt{S(S+1)}$ is the effective magnetic moment, which is temperature independent. For a spin half system with a Land\'e g-factor of 2 the effective magnetic moment is worth approximately 1.7 $\mu_\textrm{B}$. Deviations from this value can be explained by either a slightly different spin state or a g-factor different from 2, also possibly anisotropic. The fitted effective moments, $\Theta_\text{CW}$ and $\chi_0$ are given in table \ref{cwfit} and the corresponding fits can be observed in Fig. \ref{cw_fits}. The Curie Weiss temperature is isotropic within errorbars and negative, which indicate dominating antiferromagnetic interactions. The g-tensor is slightly anisotropic with easy axis along b and hard axis along a. Below 20K the susceptibility steeply increases indicating a magnetic transition into a state with a ferromagnetic component. ZFC - FC splitting at low temperature is typical of weak ferromagnetic hysteresis. The magnetic transition occurring well below $\Theta_{\textrm{CW}}$ indicates order is suppressed by fluctuations due to a combination of low dimensionality and frustration.
		\begin{figure}[htp]
			\includegraphics[width=0.45\textwidth]{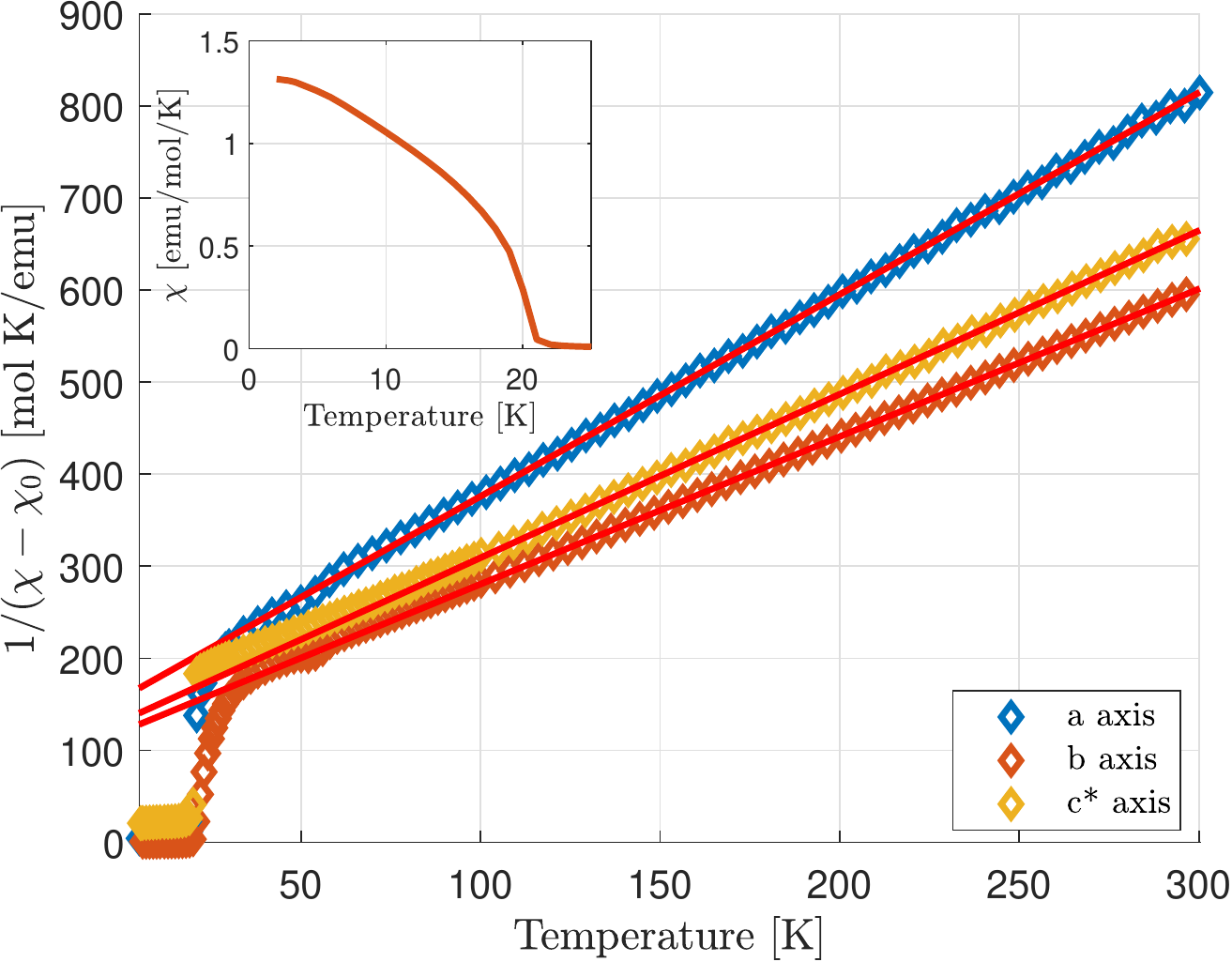}
			\caption{\label{cw_fits}Curie-Weiss fits of the high temperature susceptibility data, along the a axis(blue), b axis (red), and c* axis(yellow). The red lines indicate the fit for each direction.}
		\end{figure}
		
		\begin{table}[h]
			\begin{tabular}{|c|c|c|c|} \hline
				direction & $\Theta_{\textrm{CW}}$ [K] & $\chi_0$ & $\mu_{\textrm{eff}}$ [$\mu_B$] \\ \hline
				a &  -71(4) &     5.4e-05        & 1.91(25)\\
				b &-75(1)   &     4.7e-04        & 2.23(25)\\
				c*& -70(1)  &    -7.6e-05        & 2.11(11)\\\hline
			\end{tabular} \caption{\label{cwfit}Results of Curie-Weiss fits for measurements along a, b and c*}
		\end{table}
		
		Fig. \ref{fc_zfc} shows difference in the field cooled (FC) and zero field cooled (ZFC) DC magnetic susceptibility measured along the b axis, for different values of measurement field : H = 5 mT,  H = 10 mT, H = 20 mT,  H = 50 mT.\\
		The susceptibility along the other two crystallographic directions (not shown) (a and c*) show a similar temperature dependence, but with a much smaller value reached at low temperatures. We conclude that the susceptibility only grows along the b-direction and that the small signal in the a and c* measurements is due to a small misalignment of the crystal.
		
		\begin{figure}[htp]
			\includegraphics[width=0.45\textwidth]{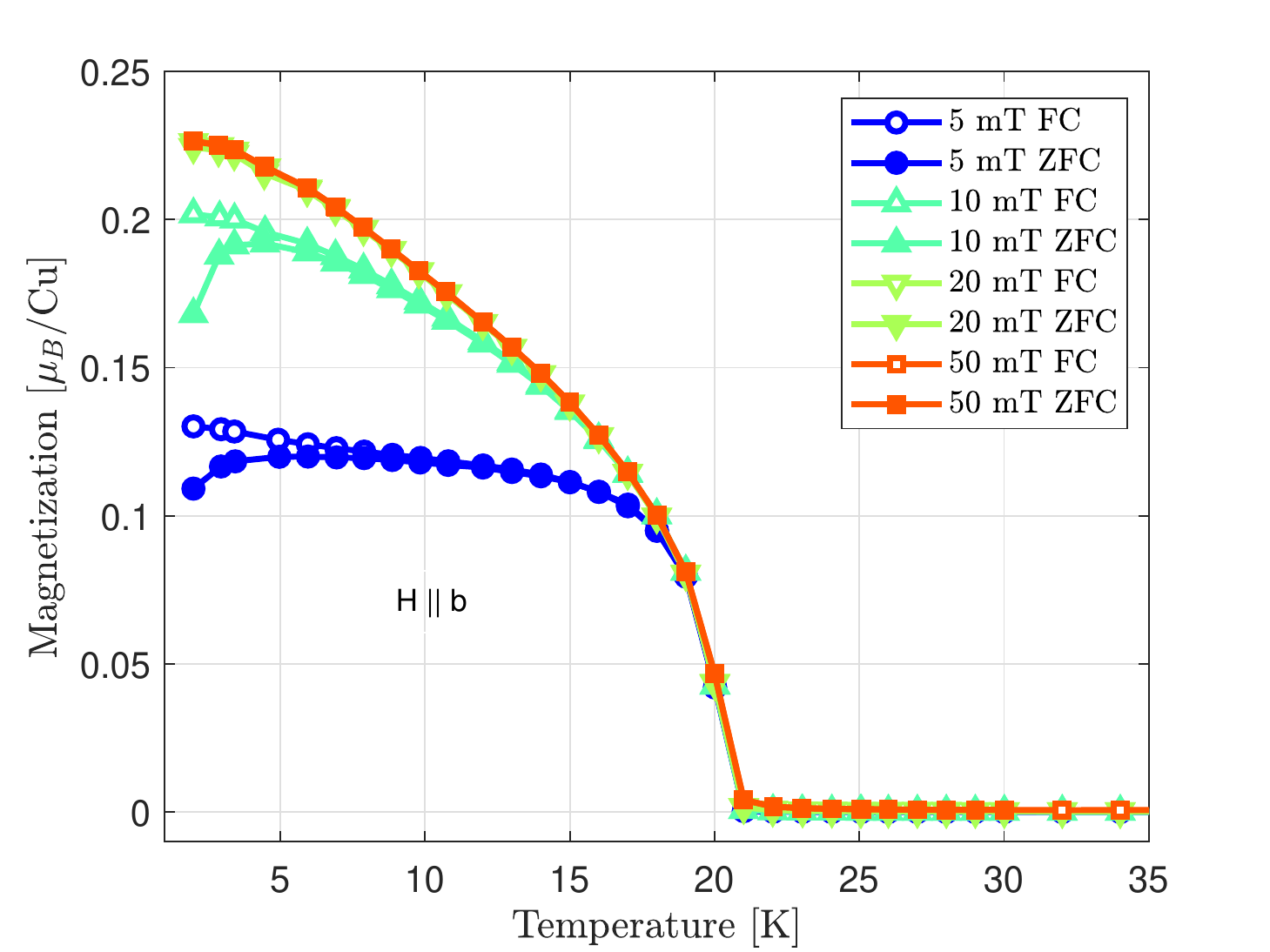}
			\caption{\label{fc_zfc} Magnetization measured on a single crystal with field along the b axis highlighting Field Cooled - Zero Field Cooled separation in low fields. The strongest separation occurs at 5 mT and decreases with field.}
		\end{figure}
		
		Fig. \ref{mh curves} shows the isothermal magnetization curves for fields parallel to the b axis. For temperatures higher than 20 K, the curves are linear, indicating a paramagnetic state. Below 20 K, a small hysteresis opens. The saturation field increases when the temperature decreases and reaches 40 mT at 2 K. No similar hysteresis opens when applying the field along the c* axis nor along the a axis (not shown).  This is another indication of a ferrimagnetic ground state with a moment $\simeq$1/4 $\mu_B$ pointing along the b axis. The hysteresis curve is slightly asymmetric, and displayed a slow time-dependence.
			This exchange-bias related phenomenon could reflect that the ferrimagnetism is the result of coupling ferromagnetic and antiferromagnetic sub-units.\\
		\begin{figure}[htp]
			\includegraphics[width=0.45\textwidth]{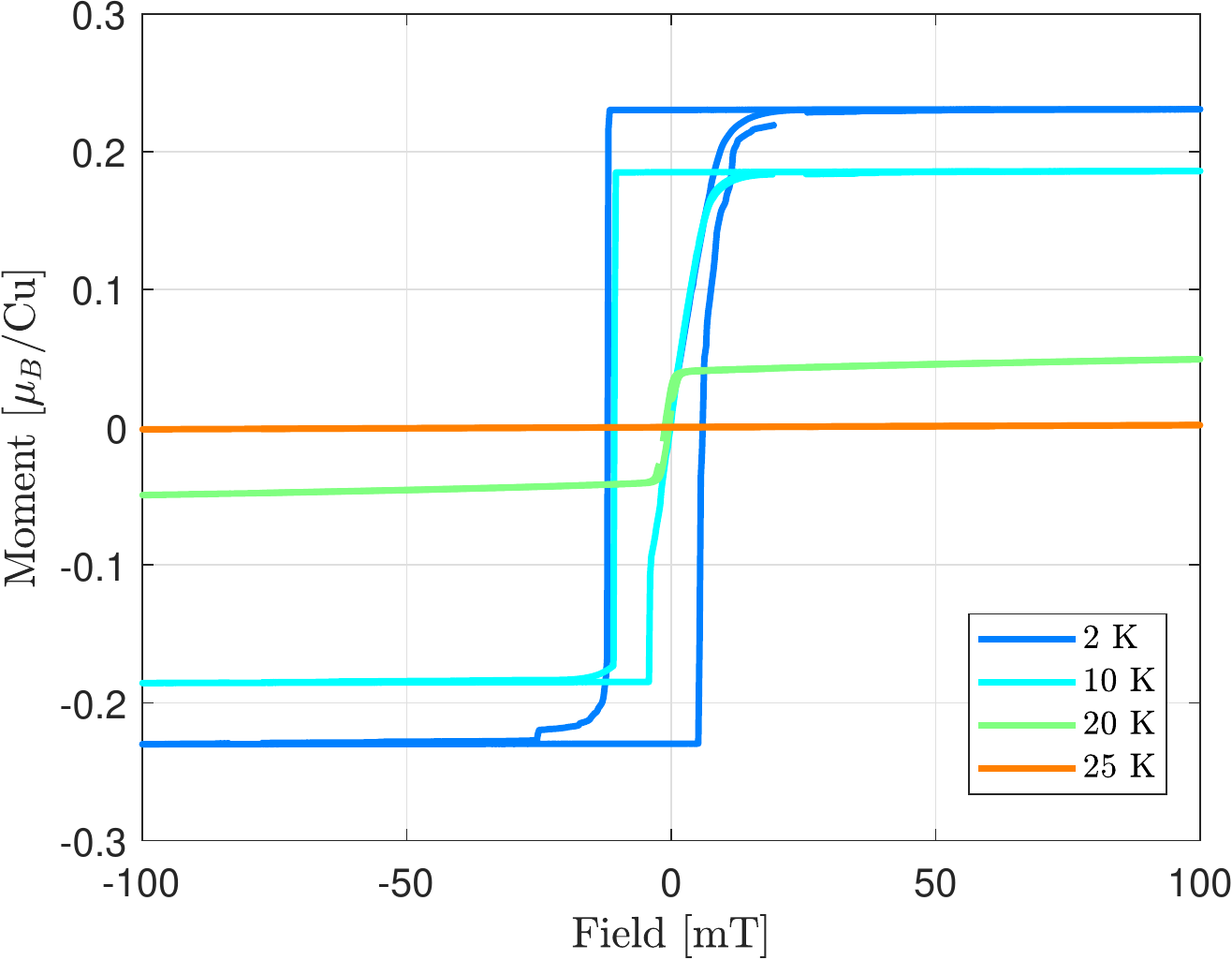}
			\caption{\label{mh curves} Isothermal magnetization for H$\left| \left| \right. \right.$b measured on a single crystal.}
		\end{figure}

		\subsection{Neutron Diffraction}
		{
%			\color{red}
			\begin{figure}[htp]
				\includegraphics[width=0.45\textwidth]{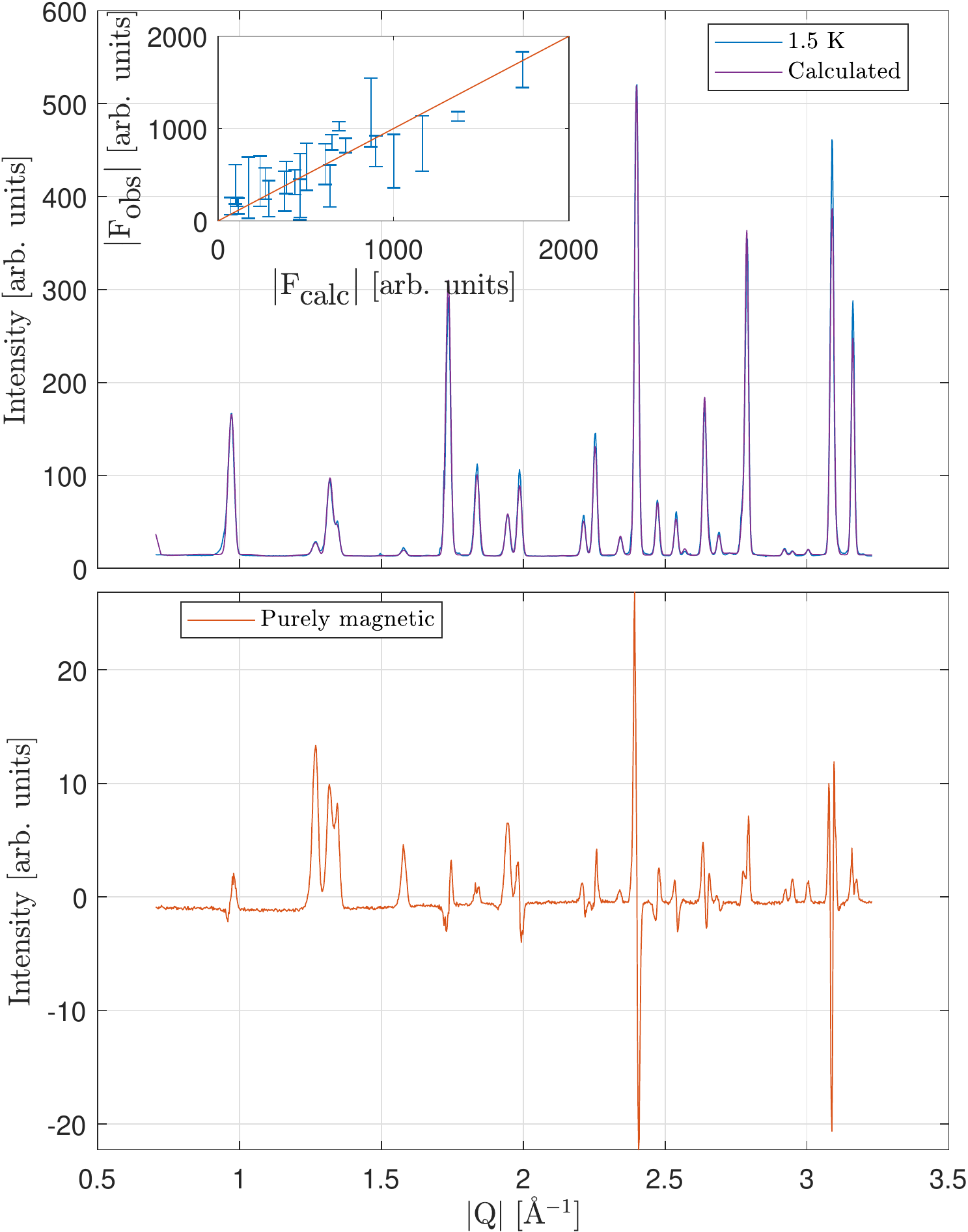}
				\caption{\label{neutron_mag_refinement} Refinement of neutron diffraction data from D20: lowest temperature data meaured at 1.5 K (blue), and result of the fit (purple). The purely magnetic signal (difference between data measured above and below the transition temperature) is shown in red. The inset show the result of magnetic refinement of the data collected at ZEBRA, with Irreps $\Gamma_1$ at a base temperature of 1.5 K. The curves displays the observed versus the calculated structure factors. The best fit shows good qualitative agreement with the model.}
			\end{figure}
			To determine the magnetic structure, neutron diffraction measurements were performed on powder samples. Fig. \ref{neutron_mag_refinement} shows in blue the data measured at 1.5 K. The magnetic scattering has been separated by measuring a powder diffraction pattern above the magnetic transition at 30 K. The subtraction of the nuclear data from the base temperature data is displayed in red in Fig. \ref{neutron_mag_refinement}. Some up-down features appear due to lattice contraction, changing the Bragg peaks positions.\\
			Every magnetic contribution is located at a nuclear Bragg peak. This indicates that the magnetic propagation vector is k=(000). Magnetic symmetry analysis was done for this propagation vector using the program BasIreps\cite{basireps, ritter2011, basir1} for the Wykoff sites 42 (Cu$_1$) and 4i (Cu$_2$) in space group C2/m. Table \ref{Irreps} lists the allowed irreducible representations (Irreps) labelled $\Gamma_1 - \Gamma_4$ and their basis vectors labelled (u,v,w) for site 1 and (r,s,t) for site 2 and shows how the magnetic moments on the symmetry related sites x,y,z and -x,y,-z are constrained and transformed by the different possible Irreps.\\
			\begin{table}
				\begin{tabular}{|c|c|c|c|c|} \hline
					~ & Cu$_1$ : x,y,z & Cu$_1$ : -x,y,-z & Cu$_2$ : x,y,z & Cu$_2$ : -x,y,-z \\ \hline
					$\Gamma_1$ & (u,v,w) & (-u,v,-w) & (0  s  0)           &  (0  s  0) \\
					$\Gamma_2$ & ~ 							   &							  ~ & (r 0 t) & (-r  0  -t)\\
					$\Gamma_3$ &(u, v, w)  & (u, -v, w) & (r 0 t) & (r 0 t) \\
					$\Gamma_4$ & ~ 							   & ~                              & (0  s  0)           &(0  -s  0) \\ \hline
				\end{tabular}
				\caption{\label{Irreps} Basis functions of irreducible representations $\Gamma_\nu$ for k = (000). Only the real components are presented because the imaginary part is zero. The two equivalent copper sites are related through the indicated transformations.}
			\end{table}
			Refinements were carried out using the Fullprof suite\cite{basireps}. The 30 K data were first refined to solve the nuclear structure. The scale factor obtained in this refinement was then fixed and used to refine the purely magnetic scattering of a dataset created by subtracting the 30 K data from the 1.5 K data having the same statistics. In doing so, the sensitivity for the magnetic contribution increases strongly reducing the uncertainties in the determination of the magnetic components.\\
			In addition to the powder diffraction we also carried out a single crystal neutron diffraction experiment on the four-circles diffractometer ZEBRA, at SINQ, PSI. The crystal used turned out to present a strong mosaicity, which did not enable us to collect intensity at Bragg reflections in a systematic way and in all the directions of reciprocal space. Intensities for a set of 55 reflections were collected at a base temperature of 1.5 K as well as at 25 K to solve the nuclear structure. The result of the fit of the experimental data is presented in the inset of Fig. \ref{neutron_mag_refinement}, for a choice of Irrep $\Gamma_1$ on both copper sites.\\
			The best refinement is given by $\Gamma_1$ on both inequivalent sites. Fig. \ref{neutron_mag_refinement} shows in purple the result of the calculation of the base temperature data using the fit result. Fig. \ref{ground state} shows the associated configuration of magnetic moments. The result of the fit of irreducible representation $\Gamma_1$ on both sites is presented in table \ref{neutron_mag_refinement_table_inplane}. If the Irreps are set to the result obtained by the powder diffraction experiment, then the free parameters in the single crystal refinement from Zebra can be reasonably adjusted and lead to a magnetic structure showing good agreement with the one obtained from the refinement of the powder data. The magnetic structure consists of Cu$_1$ antiferromagnetic chains with non colinear moments. This is the configuration that a 1D chain would adopt in magnetic field along the chain, with the field strength equal to half the saturation field. The chains are connected by Cu$_2$ dimers. The Cu$_2$ atoms in the dimer are ferromagnetically coupled. The copper atoms in the chains form an approximate 120 degrees structure with 4 spins: two Cu$_2$ atoms and two Cu$_1$ atoms belonging to the Cu$_1$ chains. The magnetic moments along the a and c* directions are compensated and the magnetic structure only supports a net moment along the b axis. Using values obtained from neutron diffraction in table \ref{neutron_mag_refinement_table_inplane}, we obtain the following saturation moment along the b axis : 0.23(3) $\mu_\textrm{B}$. This moment is made up of a quarter of the sum of the moments of two Cu$_1$ and two Cu$_2$ atoms, and is in perfect agreement with the isothermal magnetization shown in Fig. \ref{mh curves}. \\
			
			\begin{figure}[htp]
				\includegraphics[width=0.45\textwidth]{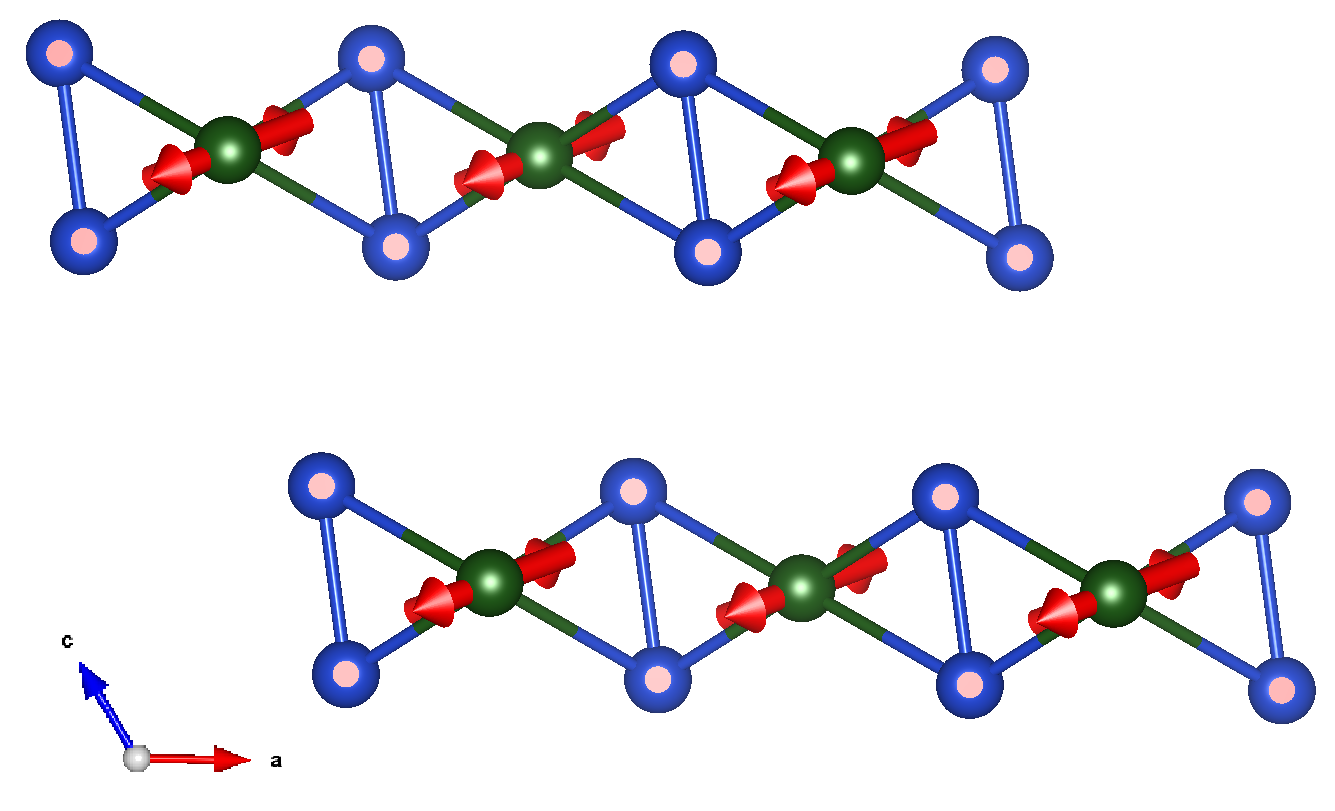}
				\includegraphics[width=0.45\textwidth]{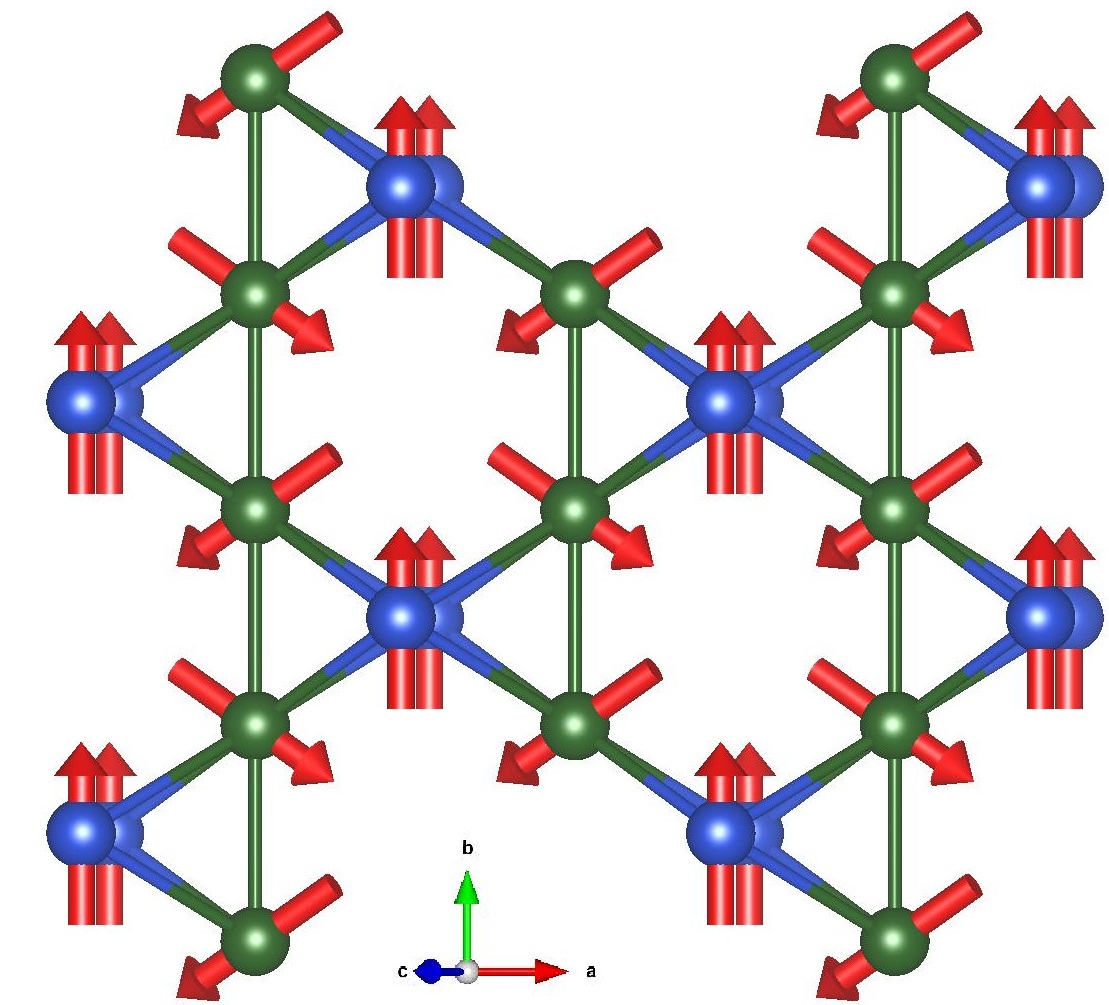}
				\caption{\label{ground state} Magnetic structure of \sample \ refined with Irrep $\Gamma_1$ on both sites. Copper site one is shown in green while copper site two is shown in blue.}
			\end{figure}

			%			\begin{table} % C:\Users\vyfavre\Documents\Work\Experiments\Cu2OSO4\D20\data\publication\magnetic from subtraction\mag_struct_sub.sum
			%				\hskip-1.25cm
			%				\begin{tabular}{|c|c|c|c|c|c|c|} \hline
			%					~ & u[$\mu_B$] & v[$\mu_B$]  & w[$\mu_B$] & s[$\mu_B$] & $\left|\left|\vec{\mu}_{\textrm{Cu}_1}\right|\right|$ [$\mu_B$]  & $\left|\left|\vec{\mu}_{\textrm{Cu}_2}\right|\right|$ [$\mu_B$] \\ \hline
			%					w $\neq 0$ & 0.83(5) & -0.47(4) &  0.30(9) & 0.86(5)   &   0.98(4) & 0.86(5)\\  \hline
			%					w = 0 & 0.70(4) & -0.45(5) &  0 & 0.91(6)&  0.83(8)   & 0.91(6) \\ \hline 
			%				\end{tabular}
			%				\caption{\label{neutron_mag_refinement_table_inplane} Result of fit to magnetic Irrep $\Gamma_1$. The second row shows results if constraining the moments to the ab-plane.}
			%			\end{table}
			
			\begin{table} % C:\Users\vyfavre\Documents\Work\Experiments\Cu2OSO4\D20\data\publication\magnetic from subtraction\mag_struct_sub.sum
				%				\hskip-1.25cm
				\begin{tabular}{|c|c|c|c|c|c|c|} \hline
					~ & u & v  & w & s & $\left|\left|\vec{\mu}_{\textrm{Cu}_1}\right|\right|$  & $\left|\left|\vec{\mu}_{\textrm{Cu}_2}\right|\right|$ \\ 
					~ & [$\mu_B$] &[$\mu_B$]  & [$\mu_B$] & [$\mu_B$] & [$\mu_B$]  & [$\mu_B$] \\ \hline
					w $\neq 0$ & 0.83(5) & -0.47(4) &  0.30(9) & 0.86(5)   &   0.98(4) & 0.86(5)\\  \hline
					w = 0 & 0.70(4) & -0.45(5) &  0 & 0.91(6)&  0.83(8)   & 0.91(6) \\ \hline 
				\end{tabular}
				\caption{\label{neutron_mag_refinement_table_inplane} Result of fit to magnetic Irrep $\Gamma_1$. The second row shows results if constraining the moments to the ab-plane.}
			\end{table}
			
			Temperature dependent neutron diffraction was performed on a powder sample. Using LBF, one can extract the associated change of the lattice parameters on the whole measured range. Fig. \ref{neutron_lattice} shows the relative change of the lattice parameters, which reflect the onset of magnetic order below 20 K.
			Fig. \ref{moments} shows the evolution of the ordered magnetic moments as a function of temperature, obtained from neutron diffraction. Solid lines correspond to power-law  fits: $\mu(T) \propto\left(T_{N}-T\right)^{2 \beta}$. The fit of the data yields $T_N$ = 19.7 (07) K, shared by the two sites, and $\beta$ = 0.12(1) for site 1 and $\beta$ = 0.13(2) for site 2.}
		
		\begin{figure}[htp]
			\includegraphics[width=0.45\textwidth]{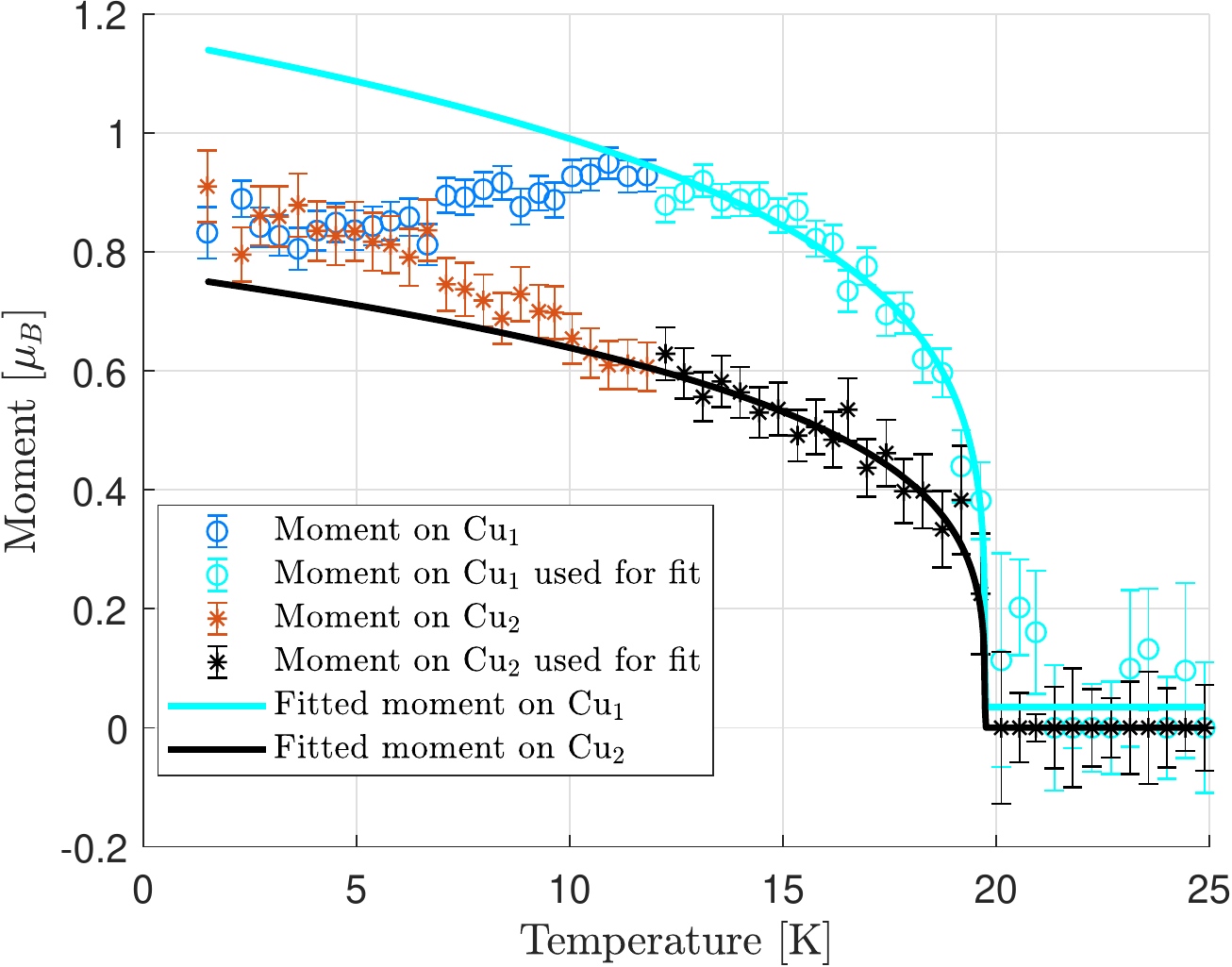}
			\caption{\label{moments} Temperature dependence of magnetic refinements with Irreps $\Gamma_1$. The solid lines are a power-law fit to the magnetic moment on each site : $\mu(T) \propto\left(T_{N}-T\right)^{2 \beta}$. The fitting of the power law behavior has been performed over the specified data range (black and cyan).}
		\end{figure}
		\section{Discussion}
%		Materials with Kagom\'e lattice of magnetic moments have been in the focus of condensed matter community for several decades now. For a long time Herbertsmithite has been the main candidate material for the realization of a quantum spin liquid state, with multiple studies converging on a conclusion that very small deviations from the ideal configuration can have detrimental effects. \textcolor{red}{In that context Cu$_2$OSO$_4$ presents a new 2D lattice topology offering an opportunity to study balance between long range order and quantum fluctuations and \st{one can regard Cu$_2$OSO$_4$ as a platform where strong deviations from a case of an ideal Kagom\'e lattice offer an opportunity to quantify those effects and provide an understanding of} their influence on a putative quantum spin liquid state.} That Cu$_2$OSO$_4$ can indeed be considered as a derivative of a \textcolor{red}{frustrated lattice \st{Kagom\'e lattice}} can be inferred from its non-zero frustration ratio, $\left|\Theta_{\textrm{CW}}\right|/T_N \sim 3.75$.\\
			
		The ordered structure determined from the neutron powder diffraction is fully consistent with exchange paths present in this compound. A close inspection of the structure reveals which are the most relevant superexchange paths. The pathway between two Cu$_2$ ions contains oxygen ions, forming the Cu$_2$ - O - Cu$_2$ angle $\sim 93.0^\circ$ which favors a ferromagnetic J$_{22}$ according to the Goodenough-Kanamori-Anderson (GKA) rules. On the other hand the Cu$_1$ - O - Cu$_1$ angle between two nearest-neighbor Cu$_1$ atoms is 114.9$^\circ$. This falls well in the range of values predicted by GKA rules to favor antiferromagnetic J$_{11}$ interactions. For the coupling between two inequivalent copper sites there are two distinct superexchange paths, one forming a Cu$_1$ - O - Cu$_2$ angle of 104.9$^\circ$ while the other one is 117.3$^\circ$. Such values are again consistent with antiferromagnetic couplings (J$_{21a}$ and J$_{21b}$, respectively).\\
		
		Additional exchange pathways exist and involve sulphur tetrahedra, more precisely Cu - O - O - Cu super-super-exchange interactions. They provide a coupling between Cu$_2$ dimers as well as an overall inter-planar coupling which eventually leads to the observed long-range order. It is a non-trivial task to estimate their relative strength to super-exchange in-plane coupling. Future inelastic neutron scattering experiments and modelling of dispersion relations could give us an indication which interactions are the most relevant for the long-range order in Cu$_2$ - O - Cu$_2$.\\
			
		As a final note, an antisymmetric Dzyaloshinskii-Moriya (DM) interaction has been indicated to exist in this compound \cite{dmCu2}, compatible with crystal symmetry. Its magnitude has been estimated to be $D \sim 7$ K. If we take $\theta_{CW} \sim 70$ K as an estimate for $J$, we arrive at $D/J \gtrsim 0.1$. For the Kagom\'e lattice, DM interactions have been predicted to cause a quantum phase transition stabilizing 120$^\circ$ order above $D_c=0.1J$\cite{Bruce, pace}. Given the similar relative order of D in \sample, it is possible that this interaction is partly responsible for the 120$^\circ$ order and as a consequence \sample \ may be close to a quantum phase transition, which could possibly be reached by tuning the system through pressure or chemical substitution.
		
		\section{Conclusion}
			In summary, we have presented a detailed study of the magnetic properties of \sample. We confirmed that the structure of \sample \ does not change upon cooling and that the ground state corresponds to a S = 1/2 system. The Curie Weiss temperatures give an idea of the order of magnitude of the magnetic couplings involved in the system, 70 K, however, the system undergoes a second order magnetic phase transition to a magnetically long range ordered state only below 20 K. Neutron scattering reveals that the ground state corresponds roughly to 120$^\circ$ order. The fact that this specific magnetic structure turns out to be the ground state, even though one third of the sites in the Kagom\'e-like lattice is replaced by ferromagnetic dimers is interesting. The rather small entropy linked to the transition suggest that there might be  interesting dynamics related to the formation of ferromagnetic pairs and 120$^\circ$ triangles and subsequent alignment of these units. We hope this study will stimulate further theoretical and experimental studies of the dynamics in this new triangular motif model compound.\\
		
		\section{Acknowledgments}
		V. Y. F.  thanks P. Babkevich for his help with the refinement of diffraction data, and F. Mila for stimulating discussions. We acknowledge the Paul Scherrer Institut, Villigen, Switzerland for provision of synchrotron radiation beamtime at beamline MS-X04SA of the SLS. This work is partially based on experiments performed at the Swiss spallation neutron source SINQ, Paul Scherrer Institute, Villigen, Switzerland, at the STFC ISIS Facility\cite{isisDOI} and at the Institut Laue-Langevin\cite{illDOI}. A portion of this research used resources at the High Flux Isotope Reactor, a DOE Office of Science User Facility operated by the Oak Ridge National Laboratory. This work was supported by the Swiss National Science Foundation (SNSF) grant No. 188648.

%\section{Reference}

%\bibliography{final} 

\end{document}